\documentclass[%
 reprint,
superscriptaddress,
 amsmath,amssymb,
 aps
]{revtex4-1}

\usepackage{verbatim}
\usepackage{graphicx}
\usepackage{dcolumn}
\usepackage{bm}
\usepackage{xcolor}
\usepackage{tipa}

\graphicspath{{./figures/}}

\begin{document}


\title{Hotter isn't faster for a melting RNA hairpin}

\author{Huaping Li}
\affiliation{Wenzhou Institute, University of Chinese Academy of Sciences, Wenzhou, Zhejiang 325001, China}
\author{E. M. Bah\c ceci}
\affiliation{Processing and Performance of Materials, Department of Mechanical Engineering, Eindhoven University of Technology, 5600 MB Eindhoven, Netherlands}
\author{M. Sayar}
\affiliation{College of Engineering, Ko\c c University, Istanbul, 34450, Turkiye\\
}
\author{A. Kabak\c c\i o\u glu}%
 \email[]{akabakcioglu@ku.edu.tr}
\affiliation{%
 College of Sciences, Ko\c c University, Istanbul, 34450, Turkiye\\
}%

\begin{abstract}
We investigate the denaturation dynamics of nucleic acids through extensive molecular dynamics simulations of a coarse-grained RNA hairpin model. In apparent contradiction with Arrhenius' law, our findings reveal that the denaturation time of RNA hairpins is a non-monotonous function of temperature for molecules longer than few persistence lengths, with an optimal temperature above the melting point, $T_m$, at which denaturation occurs fastest. This anomaly arises from the existence of two distinct pathways: ``unidirectional'' unzipping, progressing from one end to the other and favored near $T_m$, and ``bidirectional'' denaturation, where competing unzipping events initiate from both ends at higher temperatures. The two regimes manifest distinct scaling laws for the melting time \textit{vs.} length, $L$, and are separated by a crossover temperature $T_\times$, with $(T_\times-T_m) \sim L^{-1}$. The results highlight the significant role of the helical structure in the out-of-equilibrium dynamics of RNA/DNA denaturation and unveil multiple surprises in a decades-old problem.

\end{abstract}

\pacs{Valid PACS appear here}
\maketitle

Nucleic acids are self-interacting polymers that can form complex structures and assume important functional roles in transcriptional, translational, and regulatory processes in cells~\cite{tinoco1999rna,burghardt2007rna,reuter2010rnastructure,zhao2021mechanical}. When subject to a change in the solvent temperature or tension, a DNA or RNA molecule goes through a structural (coil-helix) phase transition which has been of interest since the discovery of the iconic double helix form~\cite{watson1953molecular,cocco1999statistical,kafri2000dna,neupane2016direct,Neupane2017PNAS,neupane2018measuring,al2022simulations,ter2023experimental,suma2023nonequilibrium,PhysRevE.107.054501}. Recent advances in designed self-assembly of nucleic acids (DNA origami, DNA lattices)~\cite{dey2021dna} and DNA-coated nanoparticles~\cite{samanta2022programmable,wang2015crystallization} have revived this interest.

While the equilibrium theory of the coil-helix transition is well established, the folding and melting dynamics of nucleic acids are still subject to active research with interesting twists and turns. For example, the textbook, ``zipper-like'' folding of an RNA hairpin (a nucleic acid with a palindromic sequence) upon sudden quench progresses with a rate that changes in time, despite the common wisdom~\cite{Neupane2012,marenduzzo2001dynamical}. In fact, the folding time depends on the polymer length as $\tau \sim L^{\alpha_f}$, with a nontrivial dynamical exponent $\alpha_f \simeq 1.6$~\cite{Frederickx2014}. Simulations suggest that, underlying this scaling law is an out-of-equilibrium process which is governed by the interplay between the relaxation time of single-stranded (denatured) segments and that of the rotational degrees of freedom of the duplex, imposed by the helical geometry~\cite{Manghi2016,li2018role,li2019tension}. 

The role of the helix structure during denaturation (melting) can be even more dramatic, as examplified by circular DNAs (e.g., plasmids) placed in hot water~\cite{Vinograd1968,viglasky2000early}. With the winding number of the complementary strands a topological invariant, the high-temperature equilibrium state of a plasmid features a complex coexistence of denaturation bubbles and supercoiled duplexes~\cite{kabakcioglu_pre2009,bar_pre2012,marenduzzo2017}. Consequently, the associated phase transition is much softer than that of a linear chain~\cite{kabakcioglu_pre2012}.

\begin{figure}[t!]
\includegraphics[width=0.8\linewidth]{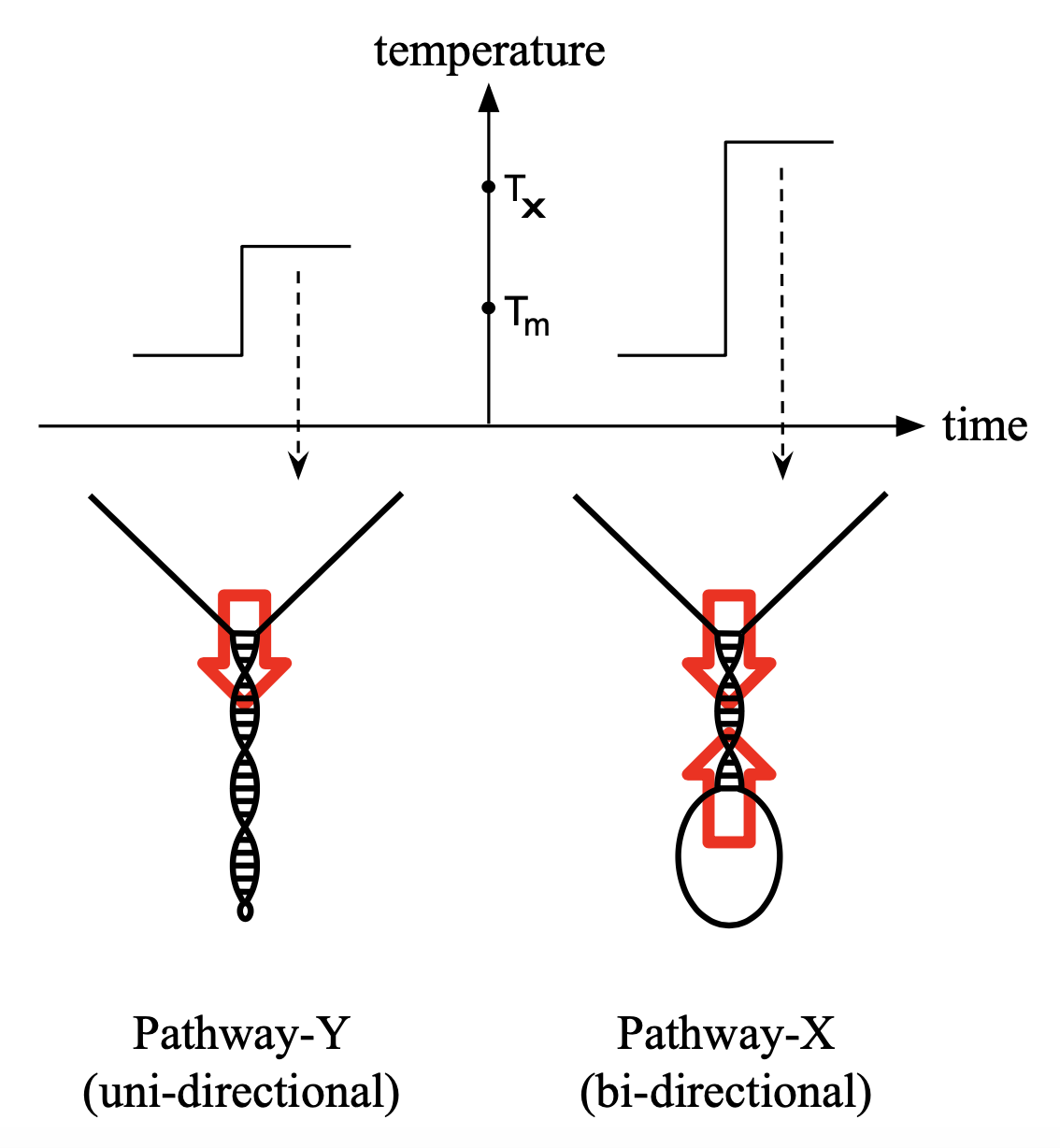}
  \caption{Schematic representation of the two distinct denaturation pathways favored below (type-Y) and above (type-X) a length-dependent crossover temperature $T_\times$ ($>T_m$), revealed by molecular dynamics simulations of a coarse-grained RNA-hairpin model.}
\label{fig_XY} 
\end{figure}

For an RNA hairpin with no such topological constraints, one might expect the melting dynamics to simply mirror the folding process in a time-reversed manner. This, too, appears not to be true: the time required for two interwined chains to thermally disentangle is reported in several studies to follow the scaling law $\tau_m \sim L^{\alpha_m}$ with $2.6\le \alpha_m \le 3.3$~\cite{baumgartner1986disentangling,Baiesi2010,baiesi2009multiple,walter2014rotational,walter2013unwinding,Walter2014,baiesi2014models}, that is, denaturation is much slower.

In this Letter, we resort to extensive molecular dynamics simulations of a coarse-grained RNA hairpin model which is subjected to a sudden temperature jump, in order to elucidate the consequences of the helicity of the bound state on the denaturation process. Our study spans more than a decade in the reduced final bath temperature, $\theta \equiv (T-T_m)/T_m \in [0.01,0.2]$, and in the chain length, $40\le L\le 640$ (measured in number of base pairs). Data unveils a new surprise: for chains longer than a few persistence lengths, we find that the mean denaturation time is a non-monotonous function of temperature! As a result, there exists an optimal solvent temperature (that depends on the length of the molecule) for which denaturation is fastest. Upon further investigation, we traced the origin of this anomaly to the presence of distinct melting pathways preferred above and below a certain crossover temperature, $T_\times$ (Fig.\ref{fig_XY}).
At relatively low temperatures ($T_m<T<T_\times$), the dominant melting pathway is the unidirectional ``unzipping'' of the duplex. For higher temperatures ($T>T_\times$) denaturation progresses from both ends, resulting in a twist-induced tug-of-war.

We will refer to these two pathways as ``Y'' ($T_m<T<T_\times$) and ``X'' ($T>T_\times$). Interestingly, despite the fact that a toy version of the type-Y ``unzipping'' process is a classic problem in undergraduate courses~\cite{kittel1980thermal}, it appears not to have been observed in experiments on or simulations of unconstrained nucleic acids so far, except for very short chains (see, e.g., Ref.\cite{Upadhyaya2021}). In fact, we show below that such unidirectional thermal denaturation is the preferred pathway only within a narrow temperature range which diminishes as $L^{-1}$.
Below, we provide numerical evidence from MD simulations that corroborates our findings summarized above. Additionally, we argue that these observations can be understood at a phenomenological level as a consequence of the twist-mediated interaction between the two ends of the duplex.

\begin{figure}[t]
\includegraphics[width=0.95\linewidth]{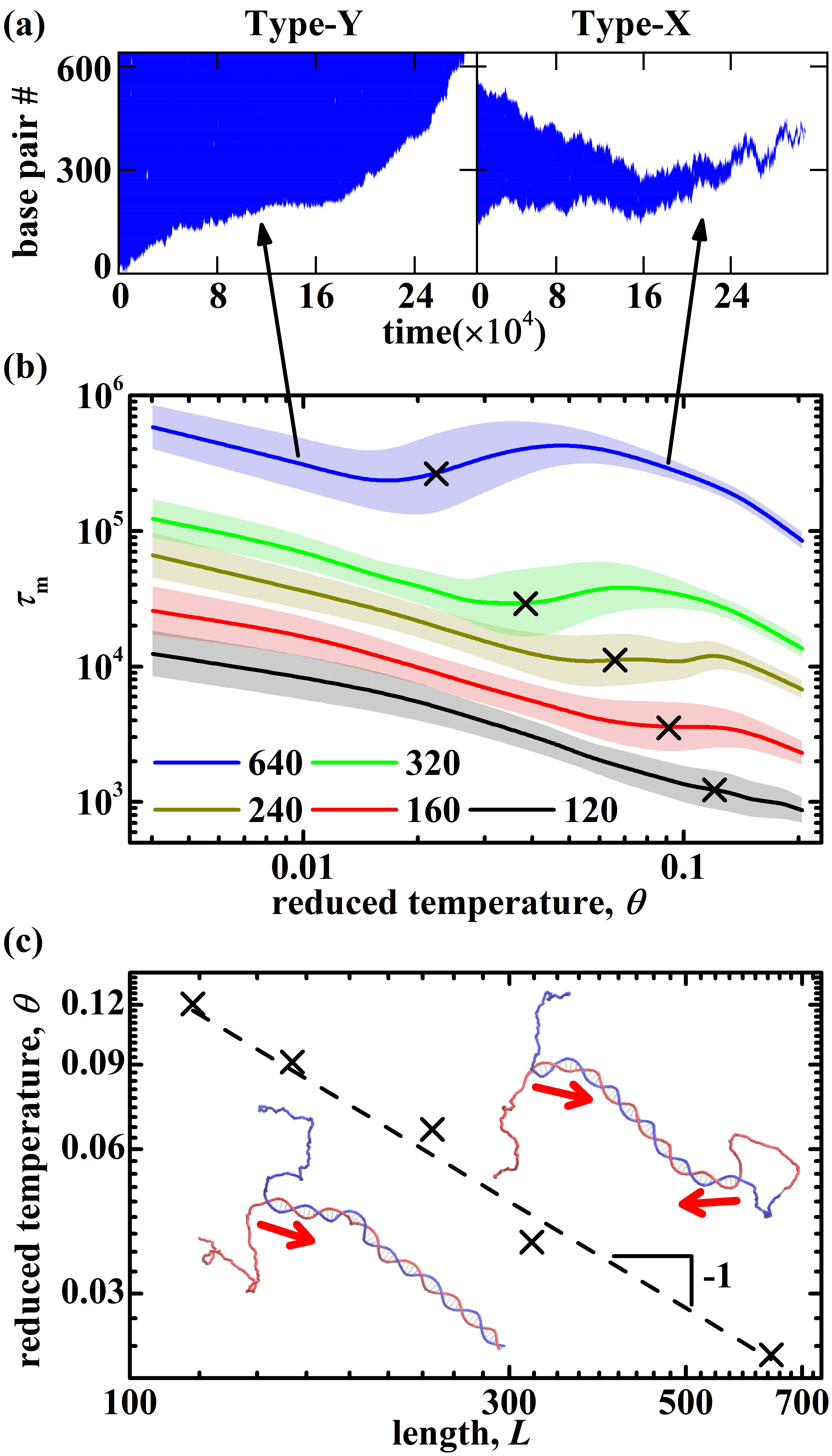}
  \caption{ (a) Representative kymographs for type-Y and type-X pathways. Vertical axis is the locus ---measured from the free end--- of bound (blue) and unbound (white) base pairs. (b) The melting time as a function of temperature for different hairpin lengths. The thickness of the shaded regions corresponds to the standard deviation around the mean (solid lines) obtained from 100 replicas.(c) The crossover temperatures, $T_\times$, shown by ``$\times$'' are defined as the position of the maximum of the coefficient of variation along each curve in (b). The dashed line is a fit with $(T_\times-T_m)/T_m \sim L^{-1}$. Two representative snapshots of type-X and type-Y denaturation, taken from the molecular dynamics simulations, are also shown.}
\label{fig:crossover}
\end{figure}

The coarse-grained MD model we used~\cite{li2018role,li2019tension} encapsulates the physical ingredients essential for this study: a helical ground state, unbreakable intra-strand and breakable inter-strand interactions, and an order of magnitude difference between the persistence lengths of the single-stranded and duplex structures. Each base is represented by a single bead in order to minimize the computation time and allow exploration of beginning-to-end denaturation processes for structures with up to $\sim 10^3$ bases. The denaturation dynamics was investigated using an RNA hairpin model embedded in three dimensions, where the complementary bases on one end  of the duplex (head) are covalently bonded while those on other end (tail) are free to dissociate. Using the in-house software developed in C++ for efficient implementation in a Langevin bath, we collected 100 independent runs at each temperature and length for sufficient statistics. For each sample, we calculated the melting time $\tau_m$ as the number of MD steps between the last instance that the fraction of broken pairs reaches $10\%$ and the first instance it reaches $90\%$. This method yields the so-called ``transition-path time'' for the denaturation process~\cite{neupane2016direct} which is dictated by the nonequilibrium dynamics of energy transfer from the nucleic acid to the environment, leaving out the somewhat arbitrary nucleation time required for the initiation of denaturation. Details of the coarse-grained model and the MD simulations can be found in the attached Supplementary Materials. 

Fig.\ref{fig:crossover} shows the temperature dependence of the measured melting times  for RNA hairpins with different lengths. The non-monotonous behavior displayed by the relatively longer structures (Fig.\ref{fig:crossover}b) is our central observation. Noting that the boiling point of water falls within $\theta\approx 0.05-0.10$ (depending on the nucleic acid composition which is ignored here), data shown in Fig.\ref{fig:crossover}b implies that for a range of hairpin lengths there exists an optimal solvent temperature at which denaturation is fastest~\footnote{A rough estimation based on Fig.\ref{fig:crossover}b places the optimum temperature 5-10 K above $T_m$ for a $500$ bps RNA hairpin, with room for variation due to sequence specificity}. This calls for a deeper exploration, since one naively expects ---by the Arrhenius law--- the pair breaking events to have a higher rate of occurrence at higher temperatures, yielding consistently shorter melting times with increasing thermal energy.

\begin{figure}[b!]
\includegraphics[width=0.95\linewidth]{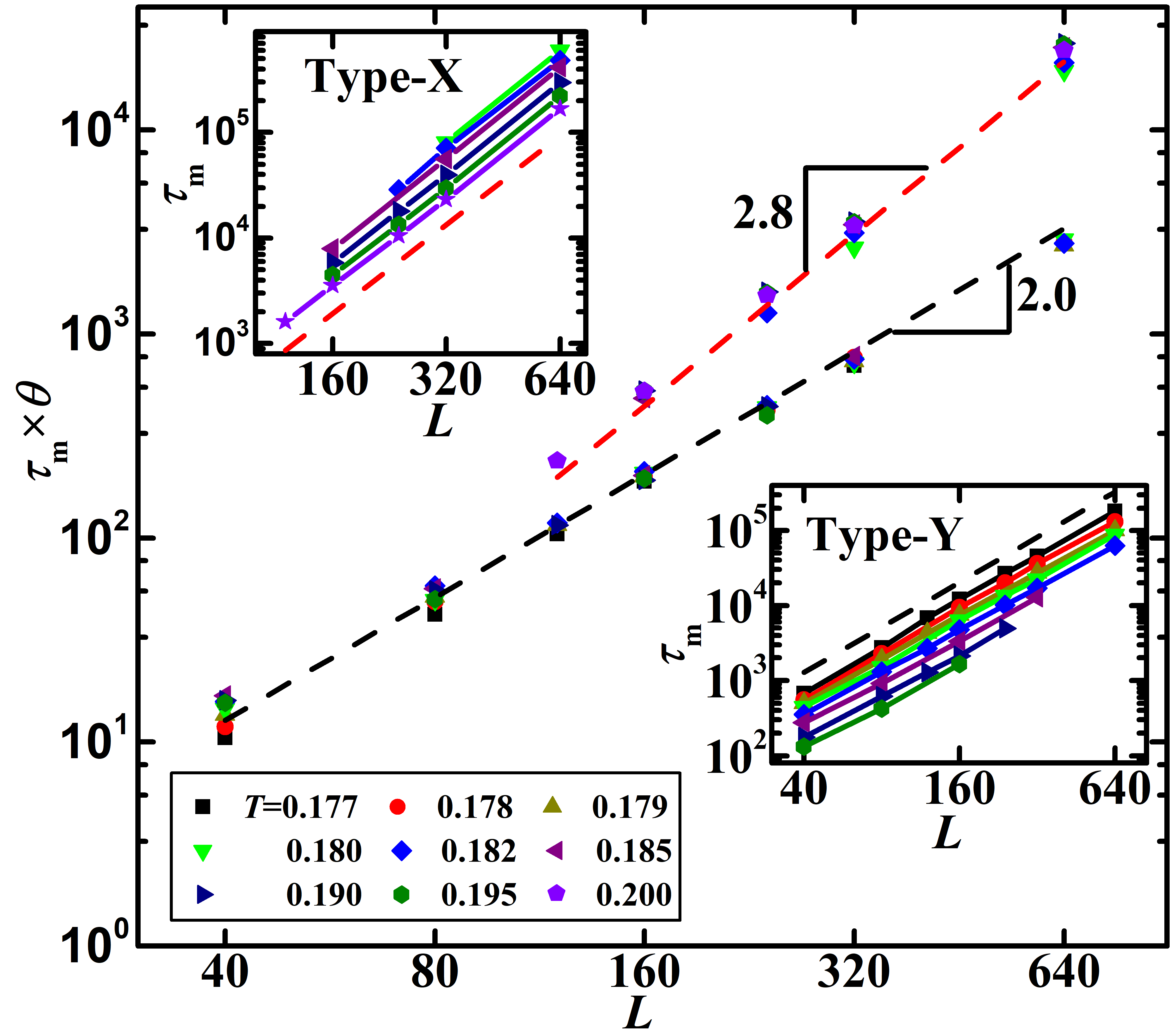}
  \caption{The melting time {\em vs.} the hairpin length. The vertical axis corresponds to the melting time scaled by the reduced temperature for data collapse, as suggested by the phenomenological arguments in the text. The slopes $2$ and $2.8$ denote the scaling exponents for type-Y and type-X melting, respectively. The raw data before temperature scaling are shown in the insets.}
  \label{fig:data_collapse}
\end{figure}

A valuable insight is gained by recognizing that  the peak of the coefficient of variation (CV) of melting times, defined as the ratio of the standard deviation to the mean, aligns with the region where non-Arrhenius behavior is observed along each curve in Fig.\ref{fig:crossover}b. 
Upon closer inspection of the individual trajectories by means of kymographs shown in Fig.\ref{fig:crossover}a, we found that the peak in CV marks the boundary between two regimes in which the denaturation process is predominantly of type ``X'' or of type ``Y'' (see Supplementary Materials). The X-Y crossover takes place at a temperature $T_\times$ which is populated by a balanced mixture of the trajectories of each type, resulting in a wide distribution of melting times. The dependence of $T_\times$ on the hairpin length (Fig.\ref{fig:crossover}c) suggests the scaling form  $(T_\times-T_m)/T_m \sim L^{-1}$ for which we propose a phenomenological explanation below.

We first address the counter-intuitive increase in the average melting time with temperature, observed in the crossover region. A plausible scenario is that type-X and type-Y dynamics are governed by different dynamical exponents, $\alpha_m^{X,Y}$, with $\alpha_m^X > \alpha_m^Y$. As a result, the crossover $Y\to X$  upon a temperature increase lands the system on a scaling law, $\tau_m(L)$, with a steeper slope. Consequently, for sufficiently large $L$, the crossover is accompanied by an increase in the melting time despite the rising temperature. To verify this hypothesis, we collected trajectories of type-X and type-Y at different temperatures (see Supplementary Materials for details) and calculated the mean melting times separately for each regime. Scaling the time data with the temperature-dependent factor $\theta^{-1}$ (more below) yields data collapse, as shown in Fig.\ref{fig:data_collapse}. 
The steeper slope of $\alpha_m^X\simeq 2.8$ for type-X melting compared to $\alpha_m^Y\simeq 2.0$ for type-Y is in agreement with the scenario proposed above. We argue below that the physical basis for these exponents is the rotational relaxation of the system in response to pair-breaking events.

In order to investigate the low-temperature behavior where the denaturation process is predominantly of type Y, consider a hairpin structure in a particular instant during the melting process, with $\ell_b$ bound base pairs forming the duplex and $\ell $ disassociated bases on each denatured tail, so that $L=\ell_b+\ell $. With the assumption that the duplex region and denatured tails contribute through the pair-binding potential and configurational entropy, respectively, the free energy can be expressed (in the spirit of the Poland-Scheraga model~\cite{Poland1966}) as
\begin{equation}
 F(\ell ) = -\epsilon (L-\ell ) - k_bT\ln\left(\frac{a^{\ell }}{\ell ^c}\right)
 \label{eq_free_energy}
\end{equation}
where $\epsilon$, and $a$ are  phenomenological constants that determine the melting temperature ($k_BT_m = \epsilon/\ln(a)$) and $c$ is a critical exponent.
Then, the entropic torsion $\kappa = -dF/d\ell $ at the y-junction is approximately constant up to a correction of $O(1/\ell )$. 
The response to this torsion is the rotation of either side of the y-junction, which we consider as the rate-limiting process. Comparing the rotation angle of the duplex and the denatured tails in the lab frame (see Supplementary Materials), we observed that the duplex's speedometer-cable-like rotation dominates the response except for a relatively brief initial period. We can then describe the dynamics of the y-junction as an overdamped motion with a fixed force and a varying friction coefficient as
\begin{equation}
\kappa = -\eta(\ell_b)\,\frac{d\ell_b}{dt} = \eta_0 (L-\ell )\,\frac{d\ell }{dt} \propto \theta\ ,
\label{eq_y_of_t}  
\end{equation}
where the friction coefficient for the rotational motion is proportional to the duplex length, $\eta_0$ is a constant, and the last proportionality ($\kappa \propto \theta$) follows from Eq.\ref{eq_free_energy}. Integration yields the melting time
\begin{equation}
\tau_m^Y \sim L^2/\theta
\label{eq_tauY}
\end{equation}
for type-Y denaturation, in agreement with the data collapse in Fig.\ref{fig:data_collapse}. 

\begin{figure}
\includegraphics[width=0.95\linewidth]{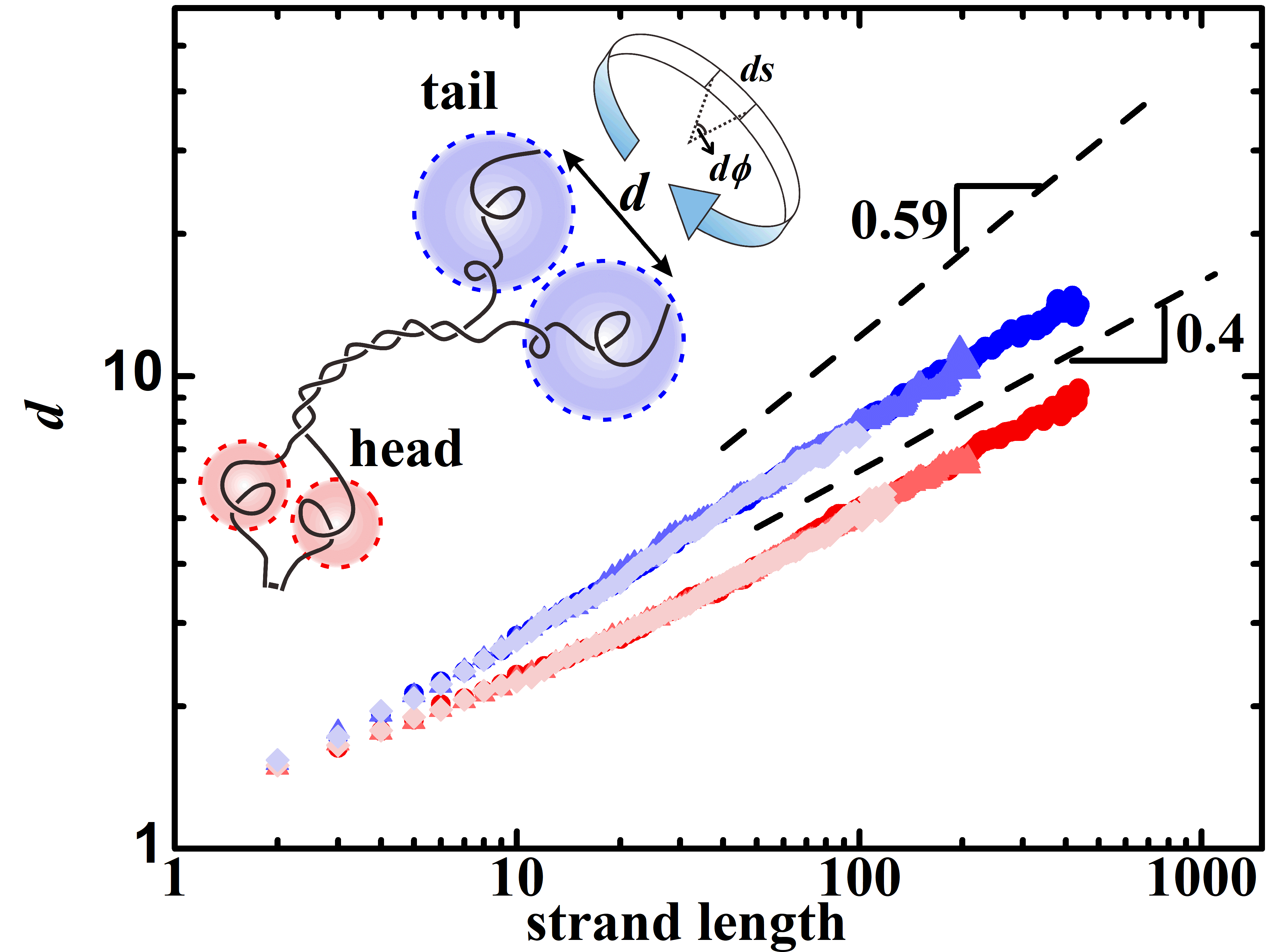}
  \caption{\label{fig:nubar} The distance between the centers of mass of denatured complementary segments as a function of segment length, with data for the head (red) and the tail (blue) of the hairpin shown separately. Darker shades correspond to data from longer hairpin models. The dashed lines are added as a guide to the eye.}
\end{figure}

    Type-X denaturation, characterized by two competing y-junctions, becomes dominant on the high-temperature side of the crossover boundary in Fig.\ref{fig:crossover}c. The duplex is now rotationally constrained, as it experiences opposing torques at both ends, and therefore cannot mediate the removal of twist from the denatured regions. As a result, the complementary denatured strands are required to rotate around each other,  leading to significantly longer melting times. To understand this regime, we adopt a phenomenological approach proposed by Baiesi {\em et al.}~\cite{Baiesi2010}, where the rotation of the denatured segments is again described by an overdamped motion governed by $dF/ds \sim \eta(\ell )\,ds/dt$. Here, $ds=\ell ^{\bar{\nu}}d\phi$ is the infinitesimal tangential displacement of the denatured strands upon rotation by an angle $d\phi$, $\bar{\nu}$ is a modified Flory exponent associated with the distance between the denatured strands and the axis of rotation, and $\eta(\ell )\sim \ell $ is the friction coefficient in the Rouse regime. Substituting, we get
\begin{equation}
    \frac{dF}{ds} = \ell ^{-\bar{\nu}}\,\frac{dF}{d\phi} \sim \ell ^{1+\bar{\nu}}\,\frac{d\phi}{dt}
\label{eq_flory_x}
\end{equation}
or, with $dF/d\phi$ approximately constant as before, 
\begin{equation}
\ell ^{1+2\bar{\nu}}\,\frac{d\phi}{dt} = const.
\label{eq_x}
\end{equation}
Observing that, in the absence of duplex rotation, the length of the denatured tails is proportional to the total unwinding angle, $\phi$, Eq.\ref{eq_x} can be integrated to yield
\begin{equation}
    \tau_m^X\sim L^{2+2\bar{\nu}}\ .
\label{eq_tauX}
\end{equation}
To test this hypothesis, we conducted a numerical investigation of the distance between complementary denatured strands along type-X trajectories. The data shown in Fig.\ref{fig:nubar} is consistent with $\bar{\nu}=0.4$, in very good agreement with Eq.\ref{eq_tauX} and $\alpha_m^X=2.8$ obtained from the data collapse in Fig.\ref{fig:data_collapse}. Notably, the divergence of $\bar{\nu}$ from the Flory exponent $\nu=0.59$ reveals that the denatured segments remain out of equilibrium during melting.

\begin{figure}[b!]
\includegraphics[width=0.95\linewidth]{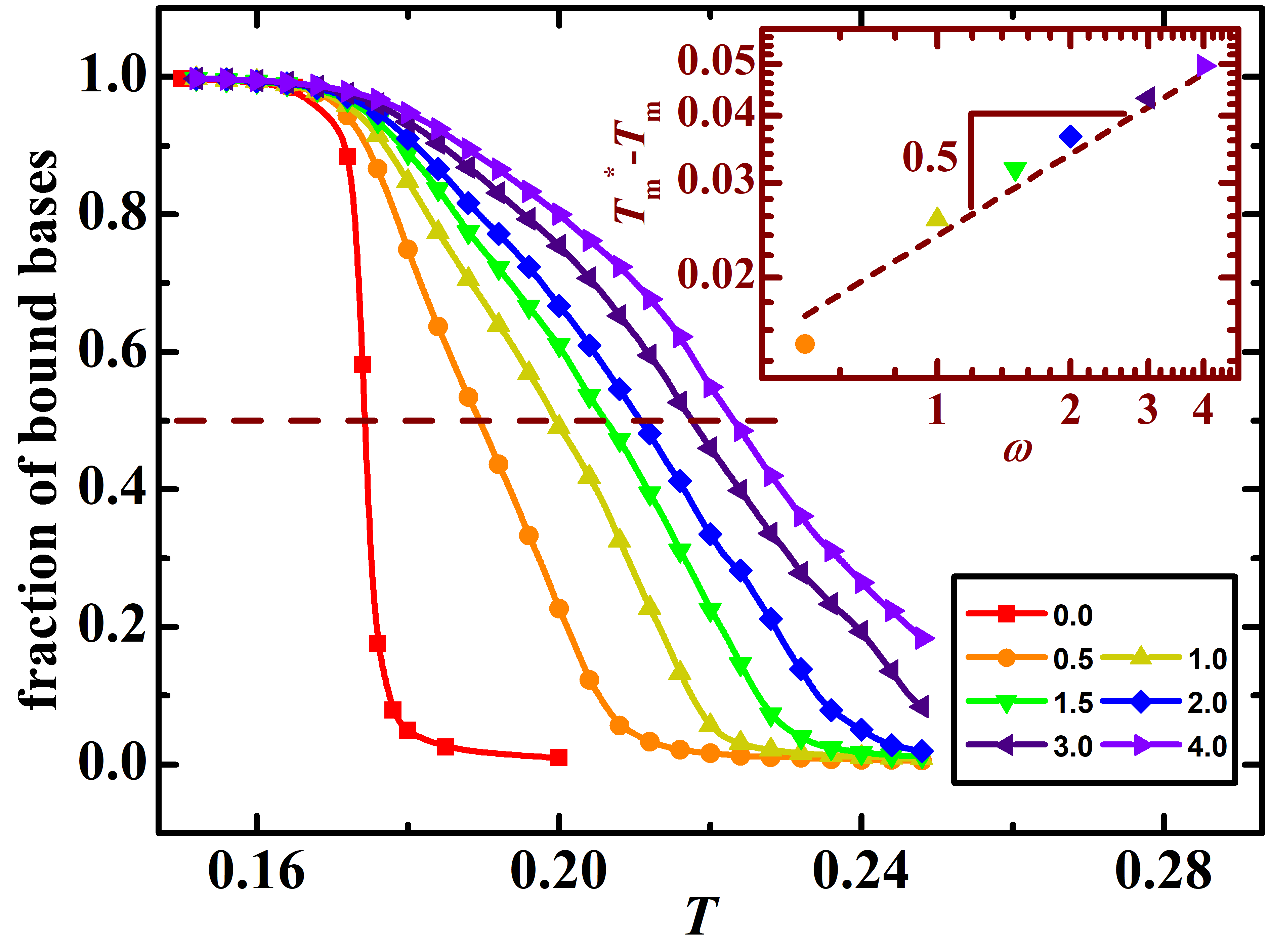}
  \caption{\label{fig6} Melting curves for an 80 bps-long hairpin model with the covalently-bonded end (head) rotated in the opposite sense of the helix. The inset shows the dependence of the melting temperature (defined by the horizontal dashed line at 50\% denaturation) on the rotation speed, measured in units of $10^{-6}$ radians per MD time step.}
\label{fig_rotate}
\end{figure}

Finally, we address our central observation in Fig.\ref{fig:crossover}: what causes the switch between the two melting behaviors as the bath temperature is changed? 
The short answer is, the speedometer-cable rotation of the duplex induced by the denaturation of one end (typically the tail, due to the entropic advantage) suppresses denaturation on the opposite end, but it can do so only if the temperature is sufficiently low. To support and quantify this claim, we performed further MD simulations of a duplex structure rotating in the opposite sense of the helix at different angular velocities and temperatures (Fig.\ref{fig_rotate}). Defining the melting point at $50\%$ denaturation ratio (as is typical for finite chains), we observed a shift in the melting temperature, $T_m^*(\omega)$, upon increasing the rotation speed, $\omega$, that conforms to
\begin{equation}
T_m^*(\omega)=T_m+a\omega^\delta
\label{eq_shift}  
\end{equation}
with $\delta=0.5$ and $a>0$ (Fig.\ref{fig_rotate} inset). 
Note that, a naive guess based on a Taylor expansion for small $\omega$ would be $\delta=1$, since we expect $T(\omega)-T_c$ to be an odd function, implying a higher thermal energy to denature upon rotation when the nucleic acid is over-wound ($\omega>0$, the present scenario) and vice versa when it is under-wound ($\omega <0$). The behavior described by Eq.7 is, therefore, quite intriguing and will be the focus of future research. A partially-bound state, similar to the intermediate temperature regime in Fig.\ref{fig_rotate}, has been observed also for a polymer tethered from one end to an impenetrable, rotating rod~\cite{laleman2016torque}.

It follows that two different melting scenarios emerge upon considering a hypothetical type-Y trajectory for a given length and solvent temperature, $T$, and requiring that $T_m^*(\omega)$ in Eq.\ref{eq_shift} remains above $T$ along the trajectory, as a consistency condition. Since $\omega$ is dictated by the speed of the y-junction, which in turn depends on the temperature  through Eq.\ref{eq_y_of_t}, type-Y denaturation dynamics turn out to be stable only in a finite temperature range.

In order to work out the X-Y crossover boundary, we substitute $ d\ell /dt \propto \omega$ in Eq.\ref{eq_y_of_t} and observe that the consistency condition above is strictest during the initial stages of denaturation, when $\omega$ is smallest and $\sim \theta/L$~\footnote{We discard the initial transient before the system settles in its long-term dynamics, during which independent denaturation events facilitated by the rotation of the denatured segments (rather than the duplex) take place at both ends.}. Consequently, the crossover temperature, $T_\times$, is found by setting  $T_\times = T^*_m(\omega)$ and substituting $\omega \sim \theta/L$ in Eq.\ref{eq_shift}:
\begin{equation}
T_\times-T_m \propto \omega(T_\times,L)^\delta \sim \left(\frac{T_\times-T_m}{L\ }\right)^\delta ,
\end{equation}
which yields
\begin{equation}
    T_\times - T_m \sim L^{-\delta/1-\delta}\ .
\label{eq_x}
\end{equation}
With $\delta=0.5$, as measured from the data in Fig.\ref{fig_rotate}, we find $T_\times \sim 1/L$, in agreement with the crossover boundary reported in Fig.\ref{fig:crossover}.

In summary, we have demonstrated through molecular dynamics simulations and phenomenological arguments that the thermal denaturation of an RNA hairpin occurs either through unidirectional unzipping progressing from one end to the other or through simultaneous unzipping of both ends, depending on the molecule size and temperature. We argued that the communication between the ends mediated by duplex rotation explains the molecule's selection between these two pathways and reproduces the observed crossover boundary. Additionally, we proposed a phenomenological framework for the dynamical scaling properties of the two denaturation regimes. Our findings reveal an unexpected twist in a decades-old problem, with predictions that should be testable using current experimental capabilities. While our numerical results are for the RNA hairpin structure with distinct boundary conditions at the two ends (Fig.~\ref{fig_XY}), the theoretical arguments presented above suggest that a DNA duplex with free boundary conditions at both ends should display a similar qualitative behavior. These observations may prove valuable in the optimization of future nanotechnological applications involving RNA/DNA self assembly.

We thank the High Performance Computing Center of Wenzhou Institute, UCAS, for the computational resources and an anonymous referee for pointing out Ref.~\cite{laleman2016torque}. This work was financially supported by the National Natural Science Foundation of China (Grant No. 22103088), Startup Fund of Wenzhou Institute, University of Chinese Academy of Sciences (Grant No. WIUCASQD2021019), and the Scientific and Technological Research Council of Turkey (Grant No. MFAG-119F354).

\bibliographystyle{apsrev4-1}
\bibliography{references}

\begin{thebibliography}{42}%
\makeatletter
\providecommand \@ifxundefined [1]{%
 \@ifx{#1\undefined}
}%
\providecommand \@ifnum [1]{%
 \ifnum #1\expandafter \@firstoftwo
 \else \expandafter \@secondoftwo
 \fi
}%
\providecommand \@ifx [1]{%
 \ifx #1\expandafter \@firstoftwo
 \else \expandafter \@secondoftwo
 \fi
}%
\providecommand \natexlab [1]{#1}%
\providecommand \enquote  [1]{``#1''}%
\providecommand \bibnamefont  [1]{#1}%
\providecommand \bibfnamefont [1]{#1}%
\providecommand \citenamefont [1]{#1}%
\providecommand \href@noop [0]{\@secondoftwo}%
\providecommand \href [0]{\begingroup \@sanitize@url \@href}%
\providecommand \@href[1]{\@@startlink{#1}\@@href}%
\providecommand \@@href[1]{\endgroup#1\@@endlink}%
\providecommand \@sanitize@url [0]{\catcode `\\12\catcode `\$12\catcode
  `\&12\catcode `\#12\catcode `\^12\catcode `\_12\catcode `\%12\relax}%
\providecommand \@@startlink[1]{}%
\providecommand \@@endlink[0]{}%
\providecommand \url  [0]{\begingroup\@sanitize@url \@url }%
\providecommand \@url [1]{\endgroup\@href {#1}{\urlprefix }}%
\providecommand \urlprefix  [0]{URL }%
\providecommand \Eprint [0]{\href }%
\providecommand \doibase [0]{http://dx.doi.org/}%
\providecommand \selectlanguage [0]{\@gobble}%
\providecommand \bibinfo  [0]{\@secondoftwo}%
\providecommand \bibfield  [0]{\@secondoftwo}%
\providecommand \translation [1]{[#1]}%
\providecommand \BibitemOpen [0]{}%
\providecommand \bibitemStop [0]{}%
\providecommand \bibitemNoStop [0]{.\EOS\space}%
\providecommand \EOS [0]{\spacefactor3000\relax}%
\providecommand \BibitemShut  [1]{\csname bibitem#1\endcsname}%
\let\auto@bib@innerbib\@empty
\bibitem [{\citenamefont {Tinoco~Jr}\ and\ \citenamefont
  {Bustamante}(1999)}]{tinoco1999rna}%
  \BibitemOpen
  \bibfield  {author} {\bibinfo {author} {\bibfnamefont {I.}~\bibnamefont
  {Tinoco~Jr}}\ and\ \bibinfo {author} {\bibfnamefont {C.}~\bibnamefont
  {Bustamante}},\ }\href@noop {} {\bibfield  {journal} {\bibinfo  {journal} {J.
  Mol. Biol.}\ }\textbf {\bibinfo {volume} {293}},\ \bibinfo {pages} {271}
  (\bibinfo {year} {1999})}\BibitemShut {NoStop}%
\bibitem [{\citenamefont {Burghardt}\ and\ \citenamefont
  {Hartmann}(2007)}]{burghardt2007rna}%
  \BibitemOpen
  \bibfield  {author} {\bibinfo {author} {\bibfnamefont {B.}~\bibnamefont
  {Burghardt}}\ and\ \bibinfo {author} {\bibfnamefont {A.~K.}\ \bibnamefont
  {Hartmann}},\ }\href@noop {} {\bibfield  {journal} {\bibinfo  {journal}
  {Phys. Rev. E}\ }\textbf {\bibinfo {volume} {75}},\ \bibinfo {pages} {021920}
  (\bibinfo {year} {2007})}\BibitemShut {NoStop}%
\bibitem [{\citenamefont {Reuter}\ and\ \citenamefont
  {Mathews}(2010)}]{reuter2010rnastructure}%
  \BibitemOpen
  \bibfield  {author} {\bibinfo {author} {\bibfnamefont {J.~S.}\ \bibnamefont
  {Reuter}}\ and\ \bibinfo {author} {\bibfnamefont {D.~H.}\ \bibnamefont
  {Mathews}},\ }\href@noop {} {\bibfield  {journal} {\bibinfo  {journal} {BMC
  Bioinformatics}\ }\textbf {\bibinfo {volume} {11}},\ \bibinfo {pages} {1}
  (\bibinfo {year} {2010})}\BibitemShut {NoStop}%
\bibitem [{\citenamefont {Zhao}\ and\ \citenamefont
  {Woodside}(2021)}]{zhao2021mechanical}%
  \BibitemOpen
  \bibfield  {author} {\bibinfo {author} {\bibfnamefont {M.}~\bibnamefont
  {Zhao}}\ and\ \bibinfo {author} {\bibfnamefont {M.~T.}\ \bibnamefont
  {Woodside}},\ }\href@noop {} {\bibfield  {journal} {\bibinfo  {journal} {Nat.
  Chem. Biol.}\ }\textbf {\bibinfo {volume} {17}},\ \bibinfo {pages} {975}
  (\bibinfo {year} {2021})}\BibitemShut {NoStop}%
\bibitem [{\citenamefont {Watson}\ and\ \citenamefont
  {Crick}(1953)}]{watson1953molecular}%
  \BibitemOpen
  \bibfield  {author} {\bibinfo {author} {\bibfnamefont {J.~D.}\ \bibnamefont
  {Watson}}\ and\ \bibinfo {author} {\bibfnamefont {F.~H.}\ \bibnamefont
  {Crick}},\ }\href@noop {} {\bibfield  {journal} {\bibinfo  {journal}
  {Nature}\ }\textbf {\bibinfo {volume} {171}},\ \bibinfo {pages} {737}
  (\bibinfo {year} {1953})}\BibitemShut {NoStop}%
\bibitem [{\citenamefont {Cocco}\ and\ \citenamefont
  {Monasson}(1999)}]{cocco1999statistical}%
  \BibitemOpen
  \bibfield  {author} {\bibinfo {author} {\bibfnamefont {S.}~\bibnamefont
  {Cocco}}\ and\ \bibinfo {author} {\bibfnamefont {R.}~\bibnamefont
  {Monasson}},\ }\href@noop {} {\bibfield  {journal} {\bibinfo  {journal}
  {Phys. Rev. Lett.}\ }\textbf {\bibinfo {volume} {83}},\ \bibinfo {pages}
  {5178} (\bibinfo {year} {1999})}\BibitemShut {NoStop}%
\bibitem [{\citenamefont {Kafri}\ \emph {et~al.}(2000)\citenamefont {Kafri},
  \citenamefont {Mukamel},\ and\ \citenamefont {Peliti}}]{kafri2000dna}%
  \BibitemOpen
  \bibfield  {author} {\bibinfo {author} {\bibfnamefont {Y.}~\bibnamefont
  {Kafri}}, \bibinfo {author} {\bibfnamefont {D.}~\bibnamefont {Mukamel}}, \
  and\ \bibinfo {author} {\bibfnamefont {L.}~\bibnamefont {Peliti}},\
  }\href@noop {} {\bibfield  {journal} {\bibinfo  {journal} {Phys. Rev. Lett.}\
  }\textbf {\bibinfo {volume} {85}},\ \bibinfo {pages} {4988} (\bibinfo {year}
  {2000})}\BibitemShut {NoStop}%
\bibitem [{\citenamefont {Neupane}\ \emph {et~al.}(2016)\citenamefont
  {Neupane}, \citenamefont {Foster}, \citenamefont {Dee}, \citenamefont {Yu},
  \citenamefont {Wang},\ and\ \citenamefont {Woodside}}]{neupane2016direct}%
  \BibitemOpen
  \bibfield  {author} {\bibinfo {author} {\bibfnamefont {K.}~\bibnamefont
  {Neupane}}, \bibinfo {author} {\bibfnamefont {D.~A.}\ \bibnamefont {Foster}},
  \bibinfo {author} {\bibfnamefont {D.~R.}\ \bibnamefont {Dee}}, \bibinfo
  {author} {\bibfnamefont {H.}~\bibnamefont {Yu}}, \bibinfo {author}
  {\bibfnamefont {F.}~\bibnamefont {Wang}}, \ and\ \bibinfo {author}
  {\bibfnamefont {M.~T.}\ \bibnamefont {Woodside}},\ }\href@noop {} {\bibfield
  {journal} {\bibinfo  {journal} {Science}\ }\textbf {\bibinfo {volume}
  {352}},\ \bibinfo {pages} {239} (\bibinfo {year} {2016})}\BibitemShut
  {NoStop}%
\bibitem [{\citenamefont {Neupane}\ \emph {et~al.}(2017)\citenamefont
  {Neupane}, \citenamefont {Wang},\ and\ \citenamefont
  {Woodside}}]{Neupane2017PNAS}%
  \BibitemOpen
  \bibfield  {author} {\bibinfo {author} {\bibfnamefont {K.}~\bibnamefont
  {Neupane}}, \bibinfo {author} {\bibfnamefont {F.}~\bibnamefont {Wang}}, \
  and\ \bibinfo {author} {\bibfnamefont {M.~T.}\ \bibnamefont {Woodside}},\
  }\href {\doibase 10.1073/pnas.1611602114} {\bibfield  {journal} {\bibinfo
  {journal} {Proc. Natl. Acad. Sci. U. S. A.}\ }\textbf {\bibinfo {volume}
  {114}},\ \bibinfo {pages} {1329} (\bibinfo {year} {2017})}\BibitemShut
  {NoStop}%
\bibitem [{\citenamefont {Neupane}\ \emph {et~al.}(2018)\citenamefont
  {Neupane}, \citenamefont {Hoffer},\ and\ \citenamefont
  {Woodside}}]{neupane2018measuring}%
  \BibitemOpen
  \bibfield  {author} {\bibinfo {author} {\bibfnamefont {K.}~\bibnamefont
  {Neupane}}, \bibinfo {author} {\bibfnamefont {N.~Q.}\ \bibnamefont {Hoffer}},
  \ and\ \bibinfo {author} {\bibfnamefont {M.}~\bibnamefont {Woodside}},\
  }\href@noop {} {\bibfield  {journal} {\bibinfo  {journal} {Phys. Rev. Lett.}\
  }\textbf {\bibinfo {volume} {121}},\ \bibinfo {pages} {018102} (\bibinfo
  {year} {2018})}\BibitemShut {NoStop}%
\bibitem [{\citenamefont {Al~Qanobi}\ \emph {et~al.}(2022)\citenamefont
  {Al~Qanobi}, \citenamefont {Marenduzzo},\ and\ \citenamefont
  {Ali}}]{al2022simulations}%
  \BibitemOpen
  \bibfield  {author} {\bibinfo {author} {\bibfnamefont {A.}~\bibnamefont
  {Al~Qanobi}}, \bibinfo {author} {\bibfnamefont {D.}~\bibnamefont
  {Marenduzzo}}, \ and\ \bibinfo {author} {\bibfnamefont {I.}~\bibnamefont
  {Ali}},\ }\href@noop {} {\bibfield  {journal} {\bibinfo  {journal} {J. Phys.:
  Cond. Matt.}\ }\textbf {\bibinfo {volume} {34}},\ \bibinfo {pages} {295101}
  (\bibinfo {year} {2022})}\BibitemShut {NoStop}%
\bibitem [{\citenamefont {Ter~Burg}\ \emph {et~al.}(2023)\citenamefont
  {Ter~Burg}, \citenamefont {Rissone}, \citenamefont {Rico-Pasto},
  \citenamefont {Ritort},\ and\ \citenamefont {Wiese}}]{ter2023experimental}%
  \BibitemOpen
  \bibfield  {author} {\bibinfo {author} {\bibfnamefont {C.}~\bibnamefont
  {Ter~Burg}}, \bibinfo {author} {\bibfnamefont {P.}~\bibnamefont {Rissone}},
  \bibinfo {author} {\bibfnamefont {M.}~\bibnamefont {Rico-Pasto}}, \bibinfo
  {author} {\bibfnamefont {F.}~\bibnamefont {Ritort}}, \ and\ \bibinfo {author}
  {\bibfnamefont {K.~J.}\ \bibnamefont {Wiese}},\ }\href@noop {} {\bibfield
  {journal} {\bibinfo  {journal} {Phys. Rev. Lett.}\ }\textbf {\bibinfo
  {volume} {130}},\ \bibinfo {pages} {208401} (\bibinfo {year}
  {2023})}\BibitemShut {NoStop}%
\bibitem [{\citenamefont {Suma}\ \emph {et~al.}(2023)\citenamefont {Suma},
  \citenamefont {Carnevale},\ and\ \citenamefont
  {Micheletti}}]{suma2023nonequilibrium}%
  \BibitemOpen
  \bibfield  {author} {\bibinfo {author} {\bibfnamefont {A.}~\bibnamefont
  {Suma}}, \bibinfo {author} {\bibfnamefont {V.}~\bibnamefont {Carnevale}}, \
  and\ \bibinfo {author} {\bibfnamefont {C.}~\bibnamefont {Micheletti}},\
  }\href@noop {} {\bibfield  {journal} {\bibinfo  {journal} {Phys. Rev. Lett.}\
  }\textbf {\bibinfo {volume} {130}},\ \bibinfo {pages} {048101} (\bibinfo
  {year} {2023})}\BibitemShut {NoStop}%
\bibitem [{\citenamefont {Rudra}\ \emph {et~al.}(2023)\citenamefont {Rudra},
  \citenamefont {Chauhan}, \citenamefont {Singh},\ and\ \citenamefont
  {Kumar}}]{PhysRevE.107.054501}%
  \BibitemOpen
  \bibfield  {author} {\bibinfo {author} {\bibfnamefont {S.}~\bibnamefont
  {Rudra}}, \bibinfo {author} {\bibfnamefont {K.}~\bibnamefont {Chauhan}},
  \bibinfo {author} {\bibfnamefont {A.~R.}\ \bibnamefont {Singh}}, \ and\
  \bibinfo {author} {\bibfnamefont {S.}~\bibnamefont {Kumar}},\ }\href
  {\doibase 10.1103/PhysRevE.107.054501} {\bibfield  {journal} {\bibinfo
  {journal} {Phys. Rev. E}\ }\textbf {\bibinfo {volume} {107}},\ \bibinfo
  {pages} {054501} (\bibinfo {year} {2023})}\BibitemShut {NoStop}%
\bibitem [{\citenamefont {Dey}\ \emph {et~al.}(2021)\citenamefont {Dey},
  \citenamefont {Fan}, \citenamefont {Gothelf}, \citenamefont {Li},
  \citenamefont {Lin}, \citenamefont {Liu}, \citenamefont {Liu}, \citenamefont
  {Nijenhuis}, \citenamefont {Sacc{\`a}}, \citenamefont {Simmel} \emph
  {et~al.}}]{dey2021dna}%
  \BibitemOpen
  \bibfield  {author} {\bibinfo {author} {\bibfnamefont {S.}~\bibnamefont
  {Dey}}, \bibinfo {author} {\bibfnamefont {C.}~\bibnamefont {Fan}}, \bibinfo
  {author} {\bibfnamefont {K.~V.}\ \bibnamefont {Gothelf}}, \bibinfo {author}
  {\bibfnamefont {J.}~\bibnamefont {Li}}, \bibinfo {author} {\bibfnamefont
  {C.}~\bibnamefont {Lin}}, \bibinfo {author} {\bibfnamefont {L.}~\bibnamefont
  {Liu}}, \bibinfo {author} {\bibfnamefont {N.}~\bibnamefont {Liu}}, \bibinfo
  {author} {\bibfnamefont {M.~A.}\ \bibnamefont {Nijenhuis}}, \bibinfo {author}
  {\bibfnamefont {B.}~\bibnamefont {Sacc{\`a}}}, \bibinfo {author}
  {\bibfnamefont {F.~C.}\ \bibnamefont {Simmel}},  \emph {et~al.},\ }\href@noop
  {} {\bibfield  {journal} {\bibinfo  {journal} {Nat. Rev. Methods Primers}\
  }\textbf {\bibinfo {volume} {1}},\ \bibinfo {pages} {13} (\bibinfo {year}
  {2021})}\BibitemShut {NoStop}%
\bibitem [{\citenamefont {Samanta}\ \emph {et~al.}(2022)\citenamefont
  {Samanta}, \citenamefont {Zhou}, \citenamefont {Ebrahimi}, \citenamefont
  {Petrosko},\ and\ \citenamefont {Mirkin}}]{samanta2022programmable}%
  \BibitemOpen
  \bibfield  {author} {\bibinfo {author} {\bibfnamefont {D.}~\bibnamefont
  {Samanta}}, \bibinfo {author} {\bibfnamefont {W.}~\bibnamefont {Zhou}},
  \bibinfo {author} {\bibfnamefont {S.~B.}\ \bibnamefont {Ebrahimi}}, \bibinfo
  {author} {\bibfnamefont {S.~H.}\ \bibnamefont {Petrosko}}, \ and\ \bibinfo
  {author} {\bibfnamefont {C.~A.}\ \bibnamefont {Mirkin}},\ }\href@noop {}
  {\bibfield  {journal} {\bibinfo  {journal} {Adv. Mater.}\ }\textbf {\bibinfo
  {volume} {34}},\ \bibinfo {pages} {2107875} (\bibinfo {year}
  {2022})}\BibitemShut {NoStop}%
\bibitem [{\citenamefont {Wang}\ \emph {et~al.}(2015)\citenamefont {Wang},
  \citenamefont {Wang}, \citenamefont {Zheng}, \citenamefont {Ducrot},
  \citenamefont {Yodh}, \citenamefont {Weck},\ and\ \citenamefont
  {Pine}}]{wang2015crystallization}%
  \BibitemOpen
  \bibfield  {author} {\bibinfo {author} {\bibfnamefont {Y.}~\bibnamefont
  {Wang}}, \bibinfo {author} {\bibfnamefont {Y.}~\bibnamefont {Wang}}, \bibinfo
  {author} {\bibfnamefont {X.}~\bibnamefont {Zheng}}, \bibinfo {author}
  {\bibfnamefont {{\'E}.}~\bibnamefont {Ducrot}}, \bibinfo {author}
  {\bibfnamefont {J.~S.}\ \bibnamefont {Yodh}}, \bibinfo {author}
  {\bibfnamefont {M.}~\bibnamefont {Weck}}, \ and\ \bibinfo {author}
  {\bibfnamefont {D.~J.}\ \bibnamefont {Pine}},\ }\href@noop {} {\bibfield
  {journal} {\bibinfo  {journal} {Nat. commun.}\ }\textbf {\bibinfo {volume}
  {6}},\ \bibinfo {pages} {7253} (\bibinfo {year} {2015})}\BibitemShut
  {NoStop}%
\bibitem [{\citenamefont {Neupane}\ \emph {et~al.}(2012)\citenamefont
  {Neupane}, \citenamefont {Ritchie}, \citenamefont {Yu}, \citenamefont
  {Foster}, \citenamefont {Wang},\ and\ \citenamefont
  {Woodside}}]{Neupane2012}%
  \BibitemOpen
  \bibfield  {author} {\bibinfo {author} {\bibfnamefont {K.}~\bibnamefont
  {Neupane}}, \bibinfo {author} {\bibfnamefont {D.~B.}\ \bibnamefont
  {Ritchie}}, \bibinfo {author} {\bibfnamefont {H.}~\bibnamefont {Yu}},
  \bibinfo {author} {\bibfnamefont {D.~A.}\ \bibnamefont {Foster}}, \bibinfo
  {author} {\bibfnamefont {F.}~\bibnamefont {Wang}}, \ and\ \bibinfo {author}
  {\bibfnamefont {M.~T.}\ \bibnamefont {Woodside}},\ }\href {\doibase
  10.1103/PhysRevLett.109.068102} {\bibfield  {journal} {\bibinfo  {journal}
  {Phys. Rev. Lett.}\ }\textbf {\bibinfo {volume} {109}},\ \bibinfo {pages}
  {068102} (\bibinfo {year} {2012})}\BibitemShut {NoStop}%
\bibitem [{\citenamefont {Marenduzzo}\ \emph {et~al.}(2001)\citenamefont
  {Marenduzzo}, \citenamefont {Bhattacharjee}, \citenamefont {Maritan},
  \citenamefont {Orlandini},\ and\ \citenamefont
  {Seno}}]{marenduzzo2001dynamical}%
  \BibitemOpen
  \bibfield  {author} {\bibinfo {author} {\bibfnamefont {D.}~\bibnamefont
  {Marenduzzo}}, \bibinfo {author} {\bibfnamefont {S.~M.}\ \bibnamefont
  {Bhattacharjee}}, \bibinfo {author} {\bibfnamefont {A.}~\bibnamefont
  {Maritan}}, \bibinfo {author} {\bibfnamefont {E.}~\bibnamefont {Orlandini}},
  \ and\ \bibinfo {author} {\bibfnamefont {F.}~\bibnamefont {Seno}},\
  }\href@noop {} {\bibfield  {journal} {\bibinfo  {journal} {Phys. Rev. Lett.}\
  }\textbf {\bibinfo {volume} {88}},\ \bibinfo {pages} {028102} (\bibinfo
  {year} {2001})}\BibitemShut {NoStop}%
\bibitem [{\citenamefont {Frederickx}\ \emph {et~al.}(2014)\citenamefont
  {Frederickx}, \citenamefont {{in't Veld}},\ and\ \citenamefont
  {Carlon}}]{Frederickx2014}%
  \BibitemOpen
  \bibfield  {author} {\bibinfo {author} {\bibfnamefont {R.}~\bibnamefont
  {Frederickx}}, \bibinfo {author} {\bibfnamefont {T.}~\bibnamefont {{in't
  Veld}}}, \ and\ \bibinfo {author} {\bibfnamefont {E.}~\bibnamefont
  {Carlon}},\ }\href {\doibase 10.1103/PhysRevLett.112.198102} {\bibfield
  {journal} {\bibinfo  {journal} {Phys. Rev. Lett.}\ }\textbf {\bibinfo
  {volume} {112}},\ \bibinfo {pages} {198102} (\bibinfo {year}
  {2014})}\BibitemShut {NoStop}%
\bibitem [{\citenamefont {Manghi}\ and\ \citenamefont
  {Destainville}(2016)}]{Manghi2016}%
  \BibitemOpen
  \bibfield  {author} {\bibinfo {author} {\bibfnamefont {M.}~\bibnamefont
  {Manghi}}\ and\ \bibinfo {author} {\bibfnamefont {N.}~\bibnamefont
  {Destainville}},\ }\href {\doibase 10.1016/j.physrep.2016.04.001} {\bibfield
  {journal} {\bibinfo  {journal} {Phys. Rep.}\ }\textbf {\bibinfo {volume}
  {631}},\ \bibinfo {pages} {1} (\bibinfo {year} {2016})}\BibitemShut {NoStop}%
\bibitem [{\citenamefont {Li}\ and\ \citenamefont
  {Kabak{\c{c}}{\i}o{\u{g}}lu}(2018)}]{li2018role}%
  \BibitemOpen
  \bibfield  {author} {\bibinfo {author} {\bibfnamefont {H.}~\bibnamefont
  {Li}}\ and\ \bibinfo {author} {\bibfnamefont {A.}~\bibnamefont
  {Kabak{\c{c}}{\i}o{\u{g}}lu}},\ }\href@noop {} {\bibfield  {journal}
  {\bibinfo  {journal} {Phys. Rev. Lett.}\ }\textbf {\bibinfo {volume} {121}},\
  \bibinfo {pages} {138101} (\bibinfo {year} {2018})}\BibitemShut {NoStop}%
\bibitem [{\citenamefont {Li}\ and\ \citenamefont
  {Kabak{\c{c}}{\i}o{\u{g}}lu}(2019)}]{li2019tension}%
  \BibitemOpen
  \bibfield  {author} {\bibinfo {author} {\bibfnamefont {H.}~\bibnamefont
  {Li}}\ and\ \bibinfo {author} {\bibfnamefont {A.}~\bibnamefont
  {Kabak{\c{c}}{\i}o{\u{g}}lu}},\ }\href@noop {} {\bibfield  {journal}
  {\bibinfo  {journal} {J. Stat. Mech.}\ }\textbf {\bibinfo {volume} {2019}},\
  \bibinfo {pages} {114005} (\bibinfo {year} {2019})}\BibitemShut {NoStop}%
\bibitem [{\citenamefont {Vinograd}\ \emph {et~al.}(1968)\citenamefont
  {Vinograd}, \citenamefont {Lebowitz},\ and\ \citenamefont
  {Watson}}]{Vinograd1968}%
  \BibitemOpen
  \bibfield  {author} {\bibinfo {author} {\bibfnamefont {J.}~\bibnamefont
  {Vinograd}}, \bibinfo {author} {\bibfnamefont {J.}~\bibnamefont {Lebowitz}},
  \ and\ \bibinfo {author} {\bibfnamefont {R.}~\bibnamefont {Watson}},\ }\href
  {\doibase https://doi.org/10.1016/0022-2836(68)90287-8} {\bibfield  {journal}
  {\bibinfo  {journal} {J. Mol. Biol.}\ }\textbf {\bibinfo {volume} {33}},\
  \bibinfo {pages} {173} (\bibinfo {year} {1968})}\BibitemShut {NoStop}%
\bibitem [{\citenamefont {V{\'\i}glask{\`y}}\ \emph {et~al.}(2000)\citenamefont
  {V{\'\i}glask{\`y}}, \citenamefont {Antal{\'\i}k}, \citenamefont
  {Adamc{\'\i}k},\ and\ \citenamefont {Podhradsk{\`y}}}]{viglasky2000early}%
  \BibitemOpen
  \bibfield  {author} {\bibinfo {author} {\bibfnamefont {V.}~\bibnamefont
  {V{\'\i}glask{\`y}}}, \bibinfo {author} {\bibfnamefont {M.}~\bibnamefont
  {Antal{\'\i}k}}, \bibinfo {author} {\bibfnamefont {J.}~\bibnamefont
  {Adamc{\'\i}k}}, \ and\ \bibinfo {author} {\bibfnamefont {D.}~\bibnamefont
  {Podhradsk{\`y}}},\ }\href@noop {} {\bibfield  {journal} {\bibinfo  {journal}
  {Nucleic Acids Res.}\ }\textbf {\bibinfo {volume} {28}},\ \bibinfo {pages}
  {e51} (\bibinfo {year} {2000})}\BibitemShut {NoStop}%
\bibitem [{\citenamefont {Kabak{\c{c}}{\i}o{\u{g}}lu}\ \emph
  {et~al.}(2009)\citenamefont {Kabak{\c{c}}{\i}o{\u{g}}lu}, \citenamefont
  {Orlandini},\ and\ \citenamefont {Mukamel}}]{kabakcioglu_pre2009}%
  \BibitemOpen
  \bibfield  {author} {\bibinfo {author} {\bibfnamefont {A.}~\bibnamefont
  {Kabak{\c{c}}{\i}o{\u{g}}lu}}, \bibinfo {author} {\bibfnamefont
  {E.}~\bibnamefont {Orlandini}}, \ and\ \bibinfo {author} {\bibfnamefont
  {D.}~\bibnamefont {Mukamel}},\ }\href@noop {} {\bibfield  {journal} {\bibinfo
   {journal} {Phys. Rev. E}\ }\textbf {\bibinfo {volume} {80}},\ \bibinfo
  {pages} {010903} (\bibinfo {year} {2009})}\BibitemShut {NoStop}%
\bibitem [{\citenamefont {Bar}\ \emph {et~al.}(2012)\citenamefont {Bar},
  \citenamefont {Kabak{\c{c}}{\i}o{\u{g}}lu},\ and\ \citenamefont
  {Mukamel}}]{bar_pre2012}%
  \BibitemOpen
  \bibfield  {author} {\bibinfo {author} {\bibfnamefont {A.}~\bibnamefont
  {Bar}}, \bibinfo {author} {\bibfnamefont {A.}~\bibnamefont
  {Kabak{\c{c}}{\i}o{\u{g}}lu}}, \ and\ \bibinfo {author} {\bibfnamefont
  {D.}~\bibnamefont {Mukamel}},\ }\href@noop {} {\bibfield  {journal} {\bibinfo
   {journal} {Phys. Rev. E}\ }\textbf {\bibinfo {volume} {86}},\ \bibinfo
  {pages} {061904} (\bibinfo {year} {2012})}\BibitemShut {NoStop}%
\bibitem [{\citenamefont {Fosado}\ \emph {et~al.}(2017)\citenamefont {Fosado},
  \citenamefont {Michieletto},\ and\ \citenamefont
  {Marenduzzo}}]{marenduzzo2017}%
  \BibitemOpen
  \bibfield  {author} {\bibinfo {author} {\bibfnamefont {Y.~A.~G.}\
  \bibnamefont {Fosado}}, \bibinfo {author} {\bibfnamefont {D.}~\bibnamefont
  {Michieletto}}, \ and\ \bibinfo {author} {\bibfnamefont {D.}~\bibnamefont
  {Marenduzzo}},\ }\href {\doibase 10.1103/PhysRevLett.119.118002} {\bibfield
  {journal} {\bibinfo  {journal} {Phys. Rev. Lett.}\ }\textbf {\bibinfo
  {volume} {119}},\ \bibinfo {pages} {118002} (\bibinfo {year}
  {2017})}\BibitemShut {NoStop}%
\bibitem [{\citenamefont {Kabak{\c{c}}{\i}o{\u{g}}lu}\ \emph
  {et~al.}(2012)\citenamefont {Kabak{\c{c}}{\i}o{\u{g}}lu}, \citenamefont
  {Bar},\ and\ \citenamefont {Mukamel}}]{kabakcioglu_pre2012}%
  \BibitemOpen
  \bibfield  {author} {\bibinfo {author} {\bibfnamefont {A.}~\bibnamefont
  {Kabak{\c{c}}{\i}o{\u{g}}lu}}, \bibinfo {author} {\bibfnamefont
  {A.}~\bibnamefont {Bar}}, \ and\ \bibinfo {author} {\bibfnamefont
  {D.}~\bibnamefont {Mukamel}},\ }\href@noop {} {\bibfield  {journal} {\bibinfo
   {journal} {Phys. Rev. E}\ }\textbf {\bibinfo {volume} {85}},\ \bibinfo
  {pages} {051919} (\bibinfo {year} {2012})}\BibitemShut {NoStop}%
\bibitem [{\citenamefont {Baumg{\"a}rtner}\ and\ \citenamefont
  {Muthukumar}(1986)}]{baumgartner1986disentangling}%
  \BibitemOpen
  \bibfield  {author} {\bibinfo {author} {\bibfnamefont {A.}~\bibnamefont
  {Baumg{\"a}rtner}}\ and\ \bibinfo {author} {\bibfnamefont {M.}~\bibnamefont
  {Muthukumar}},\ }\href@noop {} {\bibfield  {journal} {\bibinfo  {journal} {J.
  Chem. Phys.}\ }\textbf {\bibinfo {volume} {84}},\ \bibinfo {pages} {440}
  (\bibinfo {year} {1986})}\BibitemShut {NoStop}%
\bibitem [{\citenamefont {Baiesi}\ \emph {et~al.}(2010)\citenamefont {Baiesi},
  \citenamefont {Barkema}, \citenamefont {Carlon},\ and\ \citenamefont
  {Panja}}]{Baiesi2010}%
  \BibitemOpen
  \bibfield  {author} {\bibinfo {author} {\bibfnamefont {M.}~\bibnamefont
  {Baiesi}}, \bibinfo {author} {\bibfnamefont {G.~T.}\ \bibnamefont {Barkema}},
  \bibinfo {author} {\bibfnamefont {E.}~\bibnamefont {Carlon}}, \ and\ \bibinfo
  {author} {\bibfnamefont {D.}~\bibnamefont {Panja}},\ }\href {\doibase
  10.1063/1.3505551} {\bibfield  {journal} {\bibinfo  {journal} {J. Chem.
  Phys.}\ }\textbf {\bibinfo {volume} {133}},\ \bibinfo {pages} {154907}
  (\bibinfo {year} {2010})}\BibitemShut {NoStop}%
\bibitem [{\citenamefont {Baiesi}\ and\ \citenamefont
  {Livi}(2009)}]{baiesi2009multiple}%
  \BibitemOpen
  \bibfield  {author} {\bibinfo {author} {\bibfnamefont {M.}~\bibnamefont
  {Baiesi}}\ and\ \bibinfo {author} {\bibfnamefont {R.}~\bibnamefont {Livi}},\
  }\href@noop {} {\bibfield  {journal} {\bibinfo  {journal} {J. Phys. A: Math.
  Theor.}\ }\textbf {\bibinfo {volume} {42}},\ \bibinfo {pages} {082003}
  (\bibinfo {year} {2009})}\BibitemShut {NoStop}%
\bibitem [{\citenamefont {Walter}\ \emph
  {et~al.}(2014{\natexlab{a}})\citenamefont {Walter}, \citenamefont {Laleman},
  \citenamefont {Baiesi},\ and\ \citenamefont {Carlon}}]{walter2014rotational}%
  \BibitemOpen
  \bibfield  {author} {\bibinfo {author} {\bibfnamefont {J.-C.}\ \bibnamefont
  {Walter}}, \bibinfo {author} {\bibfnamefont {M.}~\bibnamefont {Laleman}},
  \bibinfo {author} {\bibfnamefont {M.}~\bibnamefont {Baiesi}}, \ and\ \bibinfo
  {author} {\bibfnamefont {E.}~\bibnamefont {Carlon}},\ }\href@noop {}
  {\bibfield  {journal} {\bibinfo  {journal} {Eur. Phys. J. Spec. Top.}\
  }\textbf {\bibinfo {volume} {223}},\ \bibinfo {pages} {3201} (\bibinfo {year}
  {2014}{\natexlab{a}})}\BibitemShut {NoStop}%
\bibitem [{\citenamefont {Walter}\ \emph {et~al.}(2013)\citenamefont {Walter},
  \citenamefont {Baiesi}, \citenamefont {Barkema},\ and\ \citenamefont
  {Carlon}}]{walter2013unwinding}%
  \BibitemOpen
  \bibfield  {author} {\bibinfo {author} {\bibfnamefont {J.-C.}\ \bibnamefont
  {Walter}}, \bibinfo {author} {\bibfnamefont {M.}~\bibnamefont {Baiesi}},
  \bibinfo {author} {\bibfnamefont {G.}~\bibnamefont {Barkema}}, \ and\
  \bibinfo {author} {\bibfnamefont {E.}~\bibnamefont {Carlon}},\ }\href@noop {}
  {\bibfield  {journal} {\bibinfo  {journal} {Phys. Rev. Lett.}\ }\textbf
  {\bibinfo {volume} {110}},\ \bibinfo {pages} {068301} (\bibinfo {year}
  {2013})}\BibitemShut {NoStop}%
\bibitem [{\citenamefont {Walter}\ \emph
  {et~al.}(2014{\natexlab{b}})\citenamefont {Walter}, \citenamefont {Baiesi},
  \citenamefont {Carlon},\ and\ \citenamefont {Schiessel}}]{Walter2014}%
  \BibitemOpen
  \bibfield  {author} {\bibinfo {author} {\bibfnamefont {J.-C.}\ \bibnamefont
  {Walter}}, \bibinfo {author} {\bibfnamefont {M.}~\bibnamefont {Baiesi}},
  \bibinfo {author} {\bibfnamefont {E.}~\bibnamefont {Carlon}}, \ and\ \bibinfo
  {author} {\bibfnamefont {H.}~\bibnamefont {Schiessel}},\ }\href {\doibase
  10.1021/ma500635h} {\bibfield  {journal} {\bibinfo  {journal}
  {Macromolecules}\ }\textbf {\bibinfo {volume} {47}},\ \bibinfo {pages} {4840}
  (\bibinfo {year} {2014}{\natexlab{b}})}\BibitemShut {NoStop}%
\bibitem [{\citenamefont {Baiesi}\ and\ \citenamefont
  {Carlon}(2013)}]{baiesi2014models}%
  \BibitemOpen
  \bibfield  {author} {\bibinfo {author} {\bibfnamefont {M.}~\bibnamefont
  {Baiesi}}\ and\ \bibinfo {author} {\bibfnamefont {E.}~\bibnamefont
  {Carlon}},\ }\href@noop {} {\bibfield  {journal} {\bibinfo  {journal} {Markov
  Processes Relat. Fields}\ }\textbf {\bibinfo {volume} {19}},\ \bibinfo
  {pages} {569} (\bibinfo {year} {2013})}\BibitemShut {NoStop}%
\bibitem [{\citenamefont {Kittel}\ and\ \citenamefont
  {Kroemer}(1980)}]{kittel1980thermal}%
  \BibitemOpen
  \bibfield  {author} {\bibinfo {author} {\bibfnamefont {C.}~\bibnamefont
  {Kittel}}\ and\ \bibinfo {author} {\bibfnamefont {H.}~\bibnamefont
  {Kroemer}},\ }\href@noop {} {\emph {\bibinfo {title} {Thermal physics}}}\
  (\bibinfo  {publisher} {Macmillan},\ \bibinfo {year} {1980})\BibitemShut
  {NoStop}%
\bibitem [{\citenamefont {Upadhyaya}\ and\ \citenamefont
  {Kumar}(2021)}]{Upadhyaya2021}%
  \BibitemOpen
  \bibfield  {author} {\bibinfo {author} {\bibfnamefont {A.}~\bibnamefont
  {Upadhyaya}}\ and\ \bibinfo {author} {\bibfnamefont {S.}~\bibnamefont
  {Kumar}},\ }\href {\doibase 10.1103/PhysRevE.103.062411} {\bibfield
  {journal} {\bibinfo  {journal} {Phys. Rev. E}\ }\textbf {\bibinfo {volume}
  {103}},\ \bibinfo {pages} {062411} (\bibinfo {year} {2021})}\BibitemShut
  {NoStop}%
\bibitem [{Note1()}]{Note1}%
  \BibitemOpen
  \bibinfo {note} {A rough estimation based on Fig.\ref {fig:crossover}b places
  the optimum temperature 5-10 K above $T_m$ for a $500$ bps RNA hairpin, with
  room for variation due to sequence specificity}\BibitemShut {NoStop}%
\bibitem [{\citenamefont {Poland}\ and\ \citenamefont
  {Scheraga}(1966)}]{Poland1966}%
  \BibitemOpen
  \bibfield  {author} {\bibinfo {author} {\bibfnamefont {D.}~\bibnamefont
  {Poland}}\ and\ \bibinfo {author} {\bibfnamefont {H.~A.}\ \bibnamefont
  {Scheraga}},\ }\href {\doibase 10.1063/1.1727785} {\bibfield  {journal}
  {\bibinfo  {journal} {J. Chem. Phys.}\ }\textbf {\bibinfo {volume} {45}},\
  \bibinfo {pages} {1456} (\bibinfo {year} {1966})}\BibitemShut {NoStop}%
\bibitem [{\citenamefont {Laleman}\ \emph {et~al.}(2016)\citenamefont
  {Laleman}, \citenamefont {Baiesi}, \citenamefont {Belotserkovskii},
  \citenamefont {Sakaue}, \citenamefont {Walter},\ and\ \citenamefont
  {Carlon}}]{laleman2016torque}%
  \BibitemOpen
  \bibfield  {author} {\bibinfo {author} {\bibfnamefont {M.}~\bibnamefont
  {Laleman}}, \bibinfo {author} {\bibfnamefont {M.}~\bibnamefont {Baiesi}},
  \bibinfo {author} {\bibfnamefont {B.~P.}\ \bibnamefont {Belotserkovskii}},
  \bibinfo {author} {\bibfnamefont {T.}~\bibnamefont {Sakaue}}, \bibinfo
  {author} {\bibfnamefont {J.-C.}\ \bibnamefont {Walter}}, \ and\ \bibinfo
  {author} {\bibfnamefont {E.}~\bibnamefont {Carlon}},\ }\href@noop {}
  {\bibfield  {journal} {\bibinfo  {journal} {Macromolecules}\ }\textbf
  {\bibinfo {volume} {49}},\ \bibinfo {pages} {405} (\bibinfo {year}
  {2016})}\BibitemShut {NoStop}%
\bibitem [{Note2()}]{Note2}%
  \BibitemOpen
  \bibinfo {note} {We discard the initial transient before the system settles
  in its long-term dynamics, during which independent denaturation events
  facilitated by the rotation of the denatured segments (rather than the
  duplex) take place at both ends.}\BibitemShut {Stop}%
\end{thebibliography}%

\end{document}